\documentstyle[12pt]{article}
\def\baselinestretch{1.2}
\catcode`\@=11

\@addtoreset{equation}{section}
\def\ksection{\arabic{section}}

\def\@normalsize{\@setsize\normalsize{15pt}\xiipt\@xiipt
\abovedisplayskip 14pt plus3pt minus3pt%
\belowdisplayskip \abovedisplayskip
\abovedisplayshortskip  \z@ plus3pt%

\belowdisplayshortskip  7pt plus3.5pt minus0pt}

\def\small{\@setsize\small{13.6pt}\xipt\@xipt
\abovedisplayskip 16pt plus3pt minus3pt%
\belowdisplayskip \abovedisplayskip
\abovedisplayshortskip  \z@ plus3pt%

\belowdisplayshortskip  7pt plus3.5pt minus0pt
\def\@listi{\parsep 4.5pt plus 2pt minus 1pt
            \itemsep \parsep
            \topsep 9pt plus 3pt minus 3pt}}

\def\underline#1{\relax\ifmmode\@@underline#1\else
	$\@@underline{\hbox{#1}}$\relax\fi}
\@twosidetrue

\catcode`@=12

\catcode`\@=11

\def\thesection{\Roman{section}.}

\def\FERMIPUB{}

\def\FERMILABPub#1{\def\FERMIPUB{#1}}
\def\ps@headings{\def\@oddfoot{}\def\@evenfoot{}
\def\@oddhead{\hbox{}\hfill
	\makebox[.5\textwidth]{\raggedright\ignorespaces --\thepage{}--
	\hfill {\rm FERMILAB--Pub--\FERMIPUB}}}
\def\@evenhead{\@oddhead}
\def\subsectionmark##1{\markboth{##1}{}}
}

\ps@headings

\catcode`\@=12

\relax

\newcounter{appendix}

\def\appendix{\par
 \addtocounter{appendix}{1}
 \def\thesection{Appendix \Alph{appendix}:}
 \def\ksection{\Alph{appendix}}}

\newskip\humongous \humongous=0pt plus 1000pt minus 1000pt

\newif\ifdtup

\def\oldreffmt#1{\rlap{[#1]} \hbox to 2\parindent{}}

\def\figfmt#1{\rlap{Figure {#1}} \hbox to 1in{}}

\def\etal{\hbox{\it et al.}}

\def\VEV#1{\left\langle #1\right\rangle}

\def\slash#1{#1\!\!\!/\!\,\,}
\def\beq{\begin{equation}}
\def\eeq{\end{equation}}
\def\bea{\begin{eqnarray}}
\def\eea{\end{eqnarray}}
\def\half{\frac{1}{2}}

\def\bq{\begin{quote}}
\def\eq{\end{quote}}

\def\half{\frac{1}{2}}

\def \lta {\mathrel{\vcenter
     {\hbox{$<$}\nointerlineskip\hbox{$\sim$}}}}
\def \gta {\mathrel{\vcenter
     {\hbox{$>$}\nointerlineskip\hbox{$\sim$}}}}

\def \etal {{\it et al.}\ }
\relax
\begin{document}
\par \vskip .05in
\FERMILABPub{94/231--T}
\begin{titlepage}
\begin{flushright}
FERMILAB--PUB--94/231--T\\
July, 1994\\
Submitted to {\em Phys. Rev.} {\bf D}
\end{flushright}
\vfill
\begin{center}
{\large \bf $Z\rightarrow b\overline{b}$ versus
Dynamical Electroweak \\ Symmetry Breaking
Involving the Top Quark
 }

 \end{center}
  \par \vskip .1in \noindent
\begin{center}
{\bf Christopher T. Hill$^1$} and
{\bf Xinmin Zhang$^2$ }
  \par \vskip .02in \noindent
{Fermi National Accelerator Laboratory\\
P.O. Box 500, Batavia, Illinois, 60510
\footnote{ Electronic address: (internet)
hill@fnal.fnal.gov,
} }

\par \vskip .02in \noindent
{Department of Physics and Astronomy\\
Iowa State University,
Ames, IA 50011 \footnote{ Electronic address: zhang@isuhep.hep.ameslab.gov
}}
  \par \vskip .02in \noindent
\end{center}

\begin{center}{\large Abstract}\end{center}
\par \vskip .01in
\begin{quote}
In models of dynamical electroweak symmetry breaking which
sensitively involve the third generation, such as top
quark condensation, the effects of the new dynamics
can show up experimentally
in $Z\rightarrow b\overline{b}$.
We compare the sensitivity of
$Z\rightarrow b\overline{b}$ and top quark production
at the Tevatron to models of the new physics.
$Z\rightarrow b\overline{b}$ is a relatively
more sensitive probe to new
strongly coupled $U(1)$ gauge bosons, while it is
generally less sensitive a probe to new physics involving
color octet gauge bosons as is top quark production itself.
Nonetheless, to accomodate a significant
excess in $Z\rightarrow b\overline{b}$ requires choosing model
parameters that may be ruled out within run I(b) at the Tevatron.
\end{quote}
 \par \vskip .02in \noindent

\vfill
\end{titlepage}
\def\baselinestretch{1.6}
\tiny
\normalsize

\noindent
{\bf 1. Introduction}
\vskip .2in
\noindent
Is the third generation special? It contains the
only fermion with a mass of order the weak scale, the top quark.
As such, the third generation, through the top quark,
may be singled out to play a key role in
electroweak symmetry or horizontal symmetry breaking physics.
At very least, it is a
privileged spectator to that dynamics.

The possibility of a heavy top quark was
anticipated within the context
of the infrared--fixed point of the Higgs--Yukawa
coupling in the standard model, \cite{FP},
and its SUSY generalization, which appears to be phenomenologically
compelling \cite{SUSY}. These ideas are,
moreover, connected to dynamical symmetry breaking
through top quark condensation \cite{BHL}.
Some authors, attempting to accomodate the
heavy top quark in technicolor schemes, have been led
to extended technicolor involving the top quark
into strong dynamics, as well
\cite{ETC}. In a gauge version of
top condensation in the standard model,
new strong physics at the $\sim 1$ TeV scale is proposed as
an imbedding of QCD: $SU(3)\rightarrow SU(3)_1\times SU(3)_2$ \cite{HILL},
where the color assignments are generationally sensitive. Thus
the effects of  new dynamics may show up experimentally in various
channels.
Anomalous top production rates and distributions at the
Tevatron might be expected because the topgluon
production mode interferes with the single gluon mode
in $q\overline{q}$ annihilation \cite{HP}. These effects
however, can potentially
be dramatic, showing up as an anomaly in the production cross-section
immediately,
or they can be subtle, requiring many thousands of top quark pairs
to become manifest.
 In technicolor schemes similar
effects can happen through pseudo--Nambu-Goldstone modes that
are produced via glue-glue collisions, \cite{EL}.  These
latter effects are
distinguished by their general kinematic structures,
such as angular distributions observable at the Tevatron \cite{KL}
and the top production  rates may differ dramatically at the LHC
from QCD.  It is important to
realize that thus far the standard model is tested only
on energy scales ranging from $\sim 0$ GeV to the $Z$--pole.
Radiative corrections largely test the running of coupling
constants on these scales and the few
instances where violation of
decoupling occurs in fermion loops, such as the $S$ and $U$ parameters,
and the $T$  (or
equivalently the $\rho$) parameter. Top quark
production represents the first time the standard model has been
examined on a new scale of $\sim 500$ GeV, and
indeed the Tevatron sensitivity to new physics
extends up to $\sim 1 $ TeV.
The emergence of new physics is certainly not prohibited
at these scales.

Dynamical symmetry breaking schemes,
such as the topcolor model, necessarily involve the $b$--quark
into the strong dynamics at the $\sim 1 $ TeV scale, or
at very least, $b_L$, since the $SU(2)_L$ group
places $(t_L, b_L)$ into a common doublet.
New dynamical effects may become manifest in
anomalous $b\overline{b}$ production at high mass  at the
Tevatron through a single ``topgluon'' interfering with
a single gluon \cite{HP}.
Moreover, sensitive studies of the $b$ quark in electroweak
production modes may reveal the new dynamics through
radiative corrections \cite{ON}, at the Tevatron through $q\overline{q}
\rightarrow W \rightarrow \overline{t}b$, or at the
NLC through $e^+e^-
\rightarrow \gamma (Z) \rightarrow \overline{b}b$,
at high $Q^2$, where the
$tbW$ or
$bb\gamma (Z)$ vertex may receive large corrections
\cite{HP2}.

In the present note we  discuss the sensitivity
of the $Z\rightarrow b\overline{b}$ rate
(measured at LEP) to new physics, such as the
topcolor model. Our approach is somewhat parallel to that
of Chivukula, et.al. \cite{CHIV}, in a discussion of
technicolor schemes.
 This process is potentially sensitive
to new physics since it is a non-universal radiative
correction,  and may probe new
forces acting in the final state at higher energies.
While  the ratio
$ R_b = \Gamma(Z\rightarrow b\overline{b})/\Gamma(Z\rightarrow
\makebox{hadrons} ) $ is slightly
high in comparison to standard
model expectations,
the discrepancy is at present only at a level of
$2\sigma$. Moreover $b$--tagging and
the various QCD contributions to $b$'s in such processes
are nontrivial issues and a possible resolution of the rate excess puzzle
in favor of a conventional effect is  quite
possible \cite{LAN}. Nevertheless, we can inquire whether
the present situation and future
prospects are potentially sensitive to new dynamics
of this sort, and to understand what limits
on dynamical models may ultimately obtain with
increasing precision in the $Z\rightarrow b\overline{b}$ rate.
The models we consider
are ``straw person'' dynamics that we are using to theoretically
assess sensitivity to new physics, and {\em we do not seek to explain
the slight $2\sigma$ excess}.

\vskip .2in
\noindent
{\bf 2. Phenomenology of $Z\rightarrow b\overline{b}$ }
\vskip .2in
\noindent
Let us take the couplings of the $b$ quark to the $Z$
boson to be given phenomenologically by the expression:
\beq
Z^\mu (g_L^{eff}\;\overline{b}\gamma_\mu b_L +
g_R^{eff}\;\overline{b}\gamma_\mu b_R) .
\eeq
We introduce two
parameters $\kappa_L$ and
$\kappa_R$ to describe the non-universal effects in the $Z b \overline{b}$
vertex. These parameters shift the standard model
tree level couplings of the
$g_{L,R}$ to effective couplings $g_{L,R}^{eff}$:
\beq
g_L^{eff}
= g_L ( 1 + \kappa_L );\qquad
g_R^{eff} = g_R ( 1 + \kappa_R ) ,
\eeq
where:
\beq
g_L = - \frac{1}{2} + \frac{1}{3}
\sin^2 \theta_W ; \qquad
g_R = \frac{1}{3} \sin^2 \theta_W .
\eeq
Defining
$\delta \Gamma$ to be the purely non-universal correction
of the new physics beyond the standard
model to the
$Z b \overline{b}$ width, $\Gamma_{b \overline{b}}$, we
have
\bea
\frac{ \delta \Gamma }{\Gamma_{b \overline{b}} }
& \sim & 2\frac{  ~( g_L^2 ~\kappa_L + g_R^2 ~\kappa_R ) }{ g_L^2 + g_R^2 }
{}~~.
\eea
\noindent
Since $g_L^2 >> g_R^2$ and the $\kappa_R$ is expected to be at most the
same order of magnitude as
 $\kappa_L$, one has approximately
\bea
\frac{ \delta \Gamma } {\Gamma_{b \overline {b}} }
& \sim &  ~2 \; \kappa_L ~~.
\eea
\noindent
Then the $R_b$ becomes:
\beq
 R_b \sim
R_b^{SM} \left( 1 + \frac{\delta \Gamma}
 {\Gamma_{b \overline{b}}} \right) \sim R_b^{SM} \left(
1 + 2 \kappa_L \right) ,
\eeq
where the standard model value
$R_b^{SM} = 0.2157 \pm 0.0004$, includes the large top quark
contributions for
$m_t = 174 \pm 11$ GeV and
$m_H = 60 \sim 1000$ GeV.
The experimental value of $R_b$ measured in aggregate at LEP is
$R_b = 0.2192 \pm 0.0018$ \cite{LAN}, which is roughly within
$2 \sigma$ of the standard model prediction.
New physics leading to
a positive $\kappa_L$, such as a short range attractive force
between $b$ and $\overline{b}$ could improve the situation, while
a negative $\kappa_L$
at the $\sim 0.5 \%$  level
would imply a discrepancy worse than $\sim 3 \sigma$
with experiment. Thus, one can
put constraints on a new physics as emphasized
by many authors (see the recent discussion of Chivukula \cite{CHIV}).

In general $\kappa_{L,R}$ can be viewed as functions of $q^2$, where
$q$ is the
$4-$momentum of the Z boson, and at LEP, $q^2 = m_Z^2$.
Expanding $\kappa_{L,R}$ in terms of $q^2$, we have
\bea
\kappa_{L,R} & \sim & \kappa_{L,R}^0 + \kappa_{L,R}^1 ~ \frac{q^2}{\Lambda^2}
{}~~,
\eea

\noindent
where ${\kappa_{L,R}^1}/{
\Lambda^2} = { d \kappa_{L(R)} }/ {d q^2}|_{q^2=0}~ $,
and
$\Lambda$ is the new physics scale.
Gauge invariant operators describing
$\kappa_{L,R}^0$ and
 $\kappa_{L,R}^1$ can always be constructed explicitly
 in a non-linear realization of $SU_L(2) \times
U_Y(1)$ \cite{PZ}.

Indeed, the above description of new physics  in terms
of modified $d=4$ current couplings  of the $b$ quark to
the $Z$ boson
is {\em apropos} the broken phase of the
standard model. These effects would also be expected
to arise from
new effective contact terms that are $d>4$, $SU(2)_L\times U(1)$
linearly invariant operators occuring at a high energy
scale $\Lambda$ above the breaking of electroweak symmetry.
For example, if
we organize the quarks into $SU(2)_L\times
SU(2)_R$ doublets as $\psi_L=(t,b)_L$ and $\psi_R=(t,b)_R$,
then  a complete basis of operators, not including the
Higgs field (here we have a dynamical symmetry breaking
in mind, and no Higgs field explicitly occuring
at the scale $\Lambda$) which
directly  mediate the process  $Z\rightarrow b\overline{b}$
can then be written:
\bea
{\cal{O}}^1_{L,R}  =  (\overline{\psi}_{L,R}\gamma_\mu
\frac{\tau^a}{2} \psi_{L,R}) (D_\nu F^{\mu\nu})^a ;
&\qquad &
{\cal{O}}^2_{L,R}  = (\overline{\psi}_{L,R}\gamma_\mu
 \frac{\tau^a}{2} D_\nu \psi_{L,R})(F^{\mu\nu})^a ;
\nonumber \\
{\cal{O}}^3_{L,R}  =  (\overline{\psi}_{L,R}\gamma_\mu
\psi_{L,R}) (\partial_\nu F^{\mu\nu}) ;
&\qquad &
{\cal{O}}^4_{L,R}  = (\overline{\psi}_{L,R}\gamma_\mu
 D_\nu \psi_{L,R})F^{\mu\nu} ;
\eea
where $F_{\mu\nu}^a$ ( $F_{\mu\nu}$ ) is the $SU(2)$ ($U(1)$) field strength.
Here we define:
\bea
\stackrel{\rightarrow}{D}_\mu & = &\stackrel{\rightarrow}{\partial}_\mu
+ ig_1A_\mu^a\frac{\tau^a}{2} + ig_2B_\mu\frac{Y}{2} ; \qquad
D_\mu  =  \half\left( \stackrel{\rightarrow}{D}_\mu -
 \stackrel{\leftarrow}{D}_\mu
\right) .
\eea
There are many other operators that can be written, but all are reducible
to this set, or operators of lower dimension,
by use of equations of motion and algebraic
identities.
Operators such as $(\overline{\psi}\gamma_\mu D_\nu \psi)
\widetilde{F}^{\mu\nu}$ are odd in CP and are not
considered, while for example  the even CP
 operator $(\overline{\psi}\gamma^5\gamma_\mu D_\nu \psi)
\widetilde{F}^{\mu\nu}$  is reducible to the above set
using massless quark equations of motion together with various identites.

In ref.\cite{BUCH} a general list
of contact terms is provided and the
 operators ${\cal{O}}^1_{L,R}$ and ${\cal{O}}^3_{L,R}$
are already reduced to four--fermion form by use of the equation of motion
$(D^\mu F_{\mu\nu})^a = g_2 j^a_\nu $.  In the broken phase of
the theory we must modify the equation of motion to include the
mass terms of the $W$ and $Z$,
e.g., $(D^\mu F_{\mu\nu})^Z = M_Z^2 Z_\nu +
g_2 j^Z_\nu $.  Thus,
 in the broken phase ${\cal{O}}^1_{L,R}$ and ${\cal{O}}^3_{L,R}$ are
 related to a four--fermion contact term and
a correction to the current couplings  of the $b$ and $t$ quarks
to the electroweak gauge bosons $\gamma$, $Z$, and $W$.
Let us introduce the effective Lagrangian containing the
contact terms:
\bea
{\cal{L}}_{eff}(M_Z) & = & \frac{1}{\Lambda^2}\left(
c^1_{X}{\cal{O}}^1_{X} +c^2_{X}{\cal{O}}^2_{X}
 +c^3_{X}{\cal{O}}^3_{X} +c^4_{X}{\cal{O}}^4_{X} + ...
\right) ,
\eea
where we sum over $X=(L,R)$ in the above.  Note that ${\cal{L}}_{eff}$
is {\em defined here at the scale $M_Z$ and not at the scale $\Lambda$}.
At the scale  $M_Z$ it is generally necessary to introduce
the Higgs field as well, which may be only an interpolating
field for a composite state (see below).
After electroweak symmetry breaking the
amplitude involving the coupling of the $Z$ boson to
the $b\overline{b}$ pair contained in ${\cal{L}}_{eff}$ becomes:
\bea
{\cal{L}}_{eff}(M_Z) & \rightarrow & -\frac{1}{\Lambda^2}\left[\left(
\half \cos\theta_W\; c^1_{X} +\sin \theta_W\; c^3_{X}
\right) M_Z^2\; \overline{b}_{X}\gamma_\mu {b}_{X} Z^{\mu}
\right.
\nonumber \\
& &
\left. + \left(
\half \cos\theta_W\; c^2_{X} +\sin \theta_W\; c^4_{X}
\right) \overline{b}_{X}\gamma_\nu \partial_\mu{b}_{X} F_Z^{\mu\nu}
\right] ,
\eea
where $F_Z^{\mu\nu}$ is the usual $U(1)$ field strength composed of a
$Z$ field.
In the first term of the {\em rhs} of eq.(11) we have
used the equation of motion of the gauge field in the
broken phase, and this term modifies the
$Z\rightarrow \overline{b}b$ current coupling.
Integrating
the second term by parts yields
zero in the limit of vanishing $m_b$, where we use
the equation of motion for the
process $Z\rightarrow \overline{b}b$.
Thus, at this stage all of the relevant
physics effects are absorbed into the definitions
of $\kappa^1_{L, R}$:
\beq
\kappa^1_{L,R} = -\frac{1}{g_Z ~ g_{L,R}}\left[
\half c^1_{L,R} \cos\theta_W\;
 + c^3_{L,R}\sin \theta_W\;
\right] ,
\eeq
\noindent
where $g_Z= e /  \sin\theta_W \cos\theta_W$.

The momentum independent
term $\kappa_{L,R}^0$ can be generated when the standard model
 Higgs doublet field $\phi$ is explicitly included. For example,
 an operator in the following form
\beq
{\cal O}^{5}_{L,R} = i ~{\overline{\psi}_{L, R}}\gamma^\mu \psi_{L,R}
          ~  \phi^\dagger D_\mu \phi ~+ ~ h.c. ~~ ,
\eeq
\noindent
 will generate a non-vanishing $\kappa_{L,R}^0$ in the broken phase
of the standard model.  Such a term can also be induced in models
in which the top quark condenses due to new strong
interaction physics affecting
the third generation.  We will presently turn to two
possible aspects of that case.

\vskip .2in
\noindent
{\bf 3. Top Condensation Models}
\vskip .1in

In the top quark condensation models,
the third generation and $t_R$ states at a minimum
participate in a new strong interaction.
In this case top and bottom quarks can be
viewed as forming the field theoretic bound state
Higgs doublet $\phi \sim {\overline{\Psi}_L} t_R$ (and
perhaps others, such as,
 $\rho$-like vector meson). Here
one would expect that the top quark loop
shown in Fig.(1) will generate the constant piece
$\kappa_{L, R}^0$  corresponding physically to the Higgs field
containing operator of eq.(13).
We will presently make an estimate
the contributions to  $\kappa_{L, R}^0$  when there are contact
terms representing the new strong dynamics.
Here, we  will limit ourselves
to the use of a momentum independent
top quark (constituent) mass up to a cut-off $\Lambda$,
and follow the method of ref. \cite{ZHANG}
in the calculation of the top loop correction to $Z b {\overline{b}}$ vertex
in the chiral lagrangian.

We begin by considering the general strength of the
induced corrections to $Z\rightarrow b\overline{b}$ from
contact terms involving the third generation.
Let us  assume the general form of the contact term to be of the
color singlet $s$--channel form as in \cite{HP}.
The relevant part of the effective Lagrangian we will take
to be  given
in the broken phase by:
\beq
{\cal L}'
  =
 - \frac{ 1}{\Lambda^2}~
 {\overline{b}}
 \gamma_\mu b ~ \overline{t} \gamma^\mu ( g_V - g_A \gamma_5
                 ) t  ~ + ...,
\eeq
\noindent
where $g_V, ~~ g_A$, are
parameters (we follow \cite{HP} for comparison of normalizations
and we will define $g_A \sim 4 \pi \times (0.11)$ below).
To compute $\kappa_{L, R}^0$ we consider Fig.(1), and we see
that we are effectively computing the top contribution
to the $Z$ boson self-energy, $\Pi_{33}$, attached to the
$b\overline{b}$ vector current.
We  then have from Fig.(1):
\beq
\kappa_{L,R}^0  =  \frac{ g_A }{g_{L,R}} \frac{ N_c}{8 \pi^2}
                       \frac{m_t^2}{\Lambda^2}\ln
                     \left( \frac{\Lambda^2}{m_t^2} \right) ,
\eeq
where $N_c = 3$.
We see that,
depending on
the sign of $g_A$, the
$Z \rightarrow b \overline{b}$ width
 can be enhanced or decreased.
In the case of a negative $\kappa_L$, requiring $\kappa_L < 0.5 \%$ gives
that $\Lambda \gta 2.0$ TeV, for $ |g_A| \sim 4 \pi \times (0.11)$
\cite{HP}. In the case of a positive $\kappa_L$, the $R_b$ can be made to be
comparable to, and even larger than the experimental value for
a decreasing $\Lambda$.
Requiring that the
 theoretical prediction be consistent with the experimental data
within $\sim 3 \sigma$, one has $\Lambda \gta 0.8$ TeV. These limits on
$\Lambda$ are comparable to, and slightly stronger than those derived from
the top quark production cross-section \cite{HP}. We present
these results for positive $\kappa_L$ in Table I. In Table I we also
assume the light fermions participate in four--fermion
operators as in eq.(14) (as in \cite{HP}) and
compare the resulting corrections to the top production cross-section.
Since this assumes that the new physics acts universally on the light
quarks, $u$ and $d$, it is in a sense an upper limit on
the effects on top production.
Thus, in this case, where the new
physics occurs in the color singlet channel,
we see that {\em the constraint from  $Z\rightarrow b\overline{b}$
is slightly stronger than from the top quark production cross-section}.

Note that in the case of  color--octet, $s$--channel,
operators,

\beq
{\cal L}'
  =
 - \frac{ 1}{\Lambda^2}~
 {\overline{b}}
 \gamma_\mu \frac{\lambda^a}{2}b ~ \overline{t} \gamma^\mu ( g_V - g_A \gamma_5
                 )\frac{\lambda^a}{2} t  ~ + ...,
\eeq
there is no contribution from the top quark owing to the
trace over colors. There is however, a contribution to the
$q^2$ dependent term from operators of this kind when the
full $SU(2)_L$ structure is considered, and it leads to
a less significant contribution to $Z \rightarrow b {\overline{b}}$.
To discuss this class of effects let us pass over to the topcolor
model \cite{HILL} which contains new heavy color octet gauge bosons,
and which leads to terms such as eq.(16) in the effective
Lagrangian.

We now assume a minimal extension of the standard model
such that at scales $\mu \gg \Lambda$, we have the gauge group
$U(1)\times SU(2)_L \times SU(3)_1\times SU(3)_2$ \cite{HILL},
with coupling constants
(gauge fields) of $SU(3)_1\times SU(3)_2$ respectively
$h_1$ and $h_2$ ($A^A_{1\mu}$ and $A^A_{2\mu}$).
We assign quark and lepton fields to anomaly free representations
as follows:
\bea
(u,\; d)_L,\;\; (c,\; s)_L & \rightarrow & \; (2,3,1); \qquad\qquad
u_R, \; d_R, \; c_R, \; s_R,  \rightarrow  \; (1,3,1);
\nonumber \\
(\nu_e,\; e )_L,\;\; (\nu_\mu,\;\mu )_L, \;\; (\nu_\tau,\; \tau )_L
 & \rightarrow  & \; (2,1,1);
 \qquad\qquad
e_R, \; \mu_R, \; \tau_R, \;( \nu_{iR} )  \rightarrow  \; (1,1,1);
\nonumber \\
(t,\; b)_L  & \rightarrow & \; (2,1,3);\qquad
t_R, \; b_R  \rightarrow \; (1,1,3);
\eea
using the notation $(SU(2)_L, SU(3)_1, SU(3)_2)$.
We break the symmetry $SU(3)_1\times SU(3)_2
\rightarrow SU(3)_c$ at the scale $M$ by introducing
a $(1,3,\bar{3})$ scalar (Higgs) field $\Phi^{a}_{b'}$
and a VEV: $\VEV{\Phi} = \makebox{diag}(M,\; M,\; M)$.
This breaks  $SU(3)_1\times SU(3)_2$ to the massless
QCD gauge group $SU(3)_c$ with gluons, $A_\mu^A$
and a residual global $SU(3)'$ with degenerate, massive
colorons, $B_\mu^A$.
The gluon $(A_\mu^A)$
and coloron $(B_\mu^A)$ fields are then defined by:
\beq
A^A_{1\mu}  = \cos\theta A^A_\mu - \sin\theta B^A_\mu ; \qquad
A^A_{2\mu}  = \sin\theta A^A_\mu + \cos\theta B^A_\mu ;
\eeq
where:
\beq
h_1\cos\theta = g_3;\qquad  h_2\sin\theta = g_3;\qquad
\tan\theta = h_1/h_2;\qquad \frac{1}{g_3^2} = \frac{1}{h_1^2} +
 \frac{1}{h_2^2} ;
\eeq
and $g_3$ is the QCD coupling constant at $M$.
We envision $h_2 \gg h_1$ and thus $\cot\theta \gg 1$
e.g., to select the top quark direction for condensation.
The mass of the degenerate octet of colorons is given by:
\beq
M_B = \left(\sqrt{h_1^2 + h_2^2}\right) M
= \left(\frac{g_3}{\sin \theta \cos \theta}\right) M .
\eeq
The usual QCD gluonic interactions
are obtained for all quarks (including top and
bottom)
while the coloron interaction takes the form:
\bea
{\cal{L}}' & =  & -\left[ g_3\cot\theta
\left( \bar{t}\gamma_\mu \frac{\lambda^A}{2} t +
 \bar{b}\gamma_\mu \frac{\lambda^A}{2} b \right)
+ \makebox{additional terms}
\right] B^{\mu A} .
\eea
If dynamical symmetry breaking occurs, or if the coloron plays a role in
inducing a large top quark mass through near critical coupling,
then $\cot\theta$ is roughly determined, e.g., the NJL result is
$h^2 = g^2_3\cot^2\theta \gta 8\pi/3$.
The computation of the
coloron radiative corrections to the $Zb\overline{b}$ vertex
 is similar to the computation
of the penguin operators of the nonleptonic weak interactions.
The diagram of Fig.(2)
gives the usual $d=4$ renormalization effects
that are absorbed into counterterms. The amplitude
corresponding to the $d=6$ operator component
of the diagram is found to be:
\bea
\; & \; &  \frac{g_Z g_3^2\cot^2\theta}{8\pi^2 M_B^2}
\left(\frac{N^2-1}{6N}\right)\ln\left(\frac{M_B^2}{M_Z^2}\right)
\overline{b} \left( q^2\gamma_\mu -q_\mu \slash{q}\right)(g_L b_L
             + g_R  b_R ) ,
\eea
where
 $N=3$.
Note that
this result can be obtained by pinching the $B$ propagator and using
penguin anomalous dimensions from the renormalization group.
This result clearly implies a nonvanishing $\kappa^1_{L,R}$.
Setting $q^2=M_Z^2$ and using current conservation with $m_b\approx 0$,
we then have:
\bea
\frac{\delta \Gamma}{\Gamma_{b \overline{b}}}
& \sim & \frac{g_3^2 M_B^2 \cot^2\theta }{ 4M_B^2 \pi^2}
\left(\frac{N^2 - 1}{6 N}\right)
         \ln \left( \frac{M_B^2} {M_Z^2} \right).
\eea
\noindent
In Table II. we present the results for $R_b$ for various
values of the model's parameters.
The coloron radiative correction is just a strong short--range attractive
force acting as a rescattering of the $b\overline{b}$ final state. It
increases the $Z \rightarrow
b \overline{b}$ width, which makes the theoretical prediction
slightly more compatable
with the LEP data.   This is expected on general grounds from any
new physics that yields an attractive interaction between the
$b$ and $\overline{b}$ in the final state.
When using the critical value
$g_3^2 \cot^2 \theta =
{8 \pi}/{3}$, the corrections to the $Z b \overline{b}$ in the topcolor
model depends on only one parameter, the mass
of the coloron, as indicated in Table II.  We see that for
small $M_B$ the effects can be large, but that for $M_B\gta 600$ GeV
only the strongest coupling can accomodate the present $R_b$ excess.
This can be compared to the effects upon $t\overline{t}$ production
at the Tevatron \cite{HP}, where a $600$ GeV coloron would
produce about a four--fold increase in the top production
cross--section relative to QCD,
and is ruled out {\em if the CDF top mass and
cross-section are accepted }\cite{CDF}. The $800$ GeV coloron
produces a cross-section for $t\overline{t}$ production that
is about a factor of $2$ greater than QCD.  Though the situation
{\em wrt} top quark production
is in flux presently, these issues should be settled by
the completion of Run I(b).  We see, however, that top production
itself is a more sensitive probe than $Z\rightarrow b\overline{b}$
in the topcolor model.

\vskip .2in
\noindent
{\bf 4. Conclusions}
\vskip .1in

In this paper we have explored the interplay between strong dynamics
that may be associated with dynamical electroweak symmetry breaking
and the large top quark mass, and the observable $R_b$ in the
process $Z\rightarrow b\overline{b}$.   This also involves the
top production cross-section at the Tevatron, and may represent a first
window on new physics beyond the standard model.  For one, our toy
models illustrate how fluid the situation is; in some cases the
effects of top production can be small when the effects upon
$Z\rightarrow b\overline{b}$ are large (as in the case of new
strongly coupled $U(1)$
interactions), while in other situations top production
is more senstive than $Z\rightarrow b\overline{b}$  (as in the
case of topcolor).
We have found the interesting result that,
if a large  anomaly in the top
quark production cross--section
is seen by CDF or D0 at the level of $\gta 50\%$, then we can  rule out
the effects of a new strong $U(1)$ interaction
as the source because it would
give rise to very large enhancements (or suppressions) of
$Z\rightarrow b\overline{b}$, as we have seen above.
This is shown in Table I. Certainly,
if the reportedly large  cross-section at CDF \cite{CDF} continues to hold,
then this cannot be explained by a $U(1)$ boson leading to
the contact term of eq.(14) since $R_b$
forces too large a value of $\Lambda$ for such terms to have
any effect upon top production at this level.

We have also  seen that, for the case of the topcolor model
which would effectively generate new color octet contact
terms, the effects in $Z\rightarrow b\overline{b}$ are not dramatic.
For example, with
$M_B \leq 600$ GeV, which may
enhance top production too  much, the discrepancy in $R_b$
with the standard model result can be reduced only from
$\sim 2 \sigma$ to $ \sim 1 \sigma$ in the limit of
large coupling.
$\kappa_L$ in the topcolor model is positive, and
 we cannot rule
out the scheme with the present
value of $R_b$ from the LEP experiments
even for critical coupling. For $M_B \gta
800$ GeV  the enhancements are within $1\sigma$ of the
standard model.  To explain an excess of
$R_b$ at the $2\sigma$ level we require
$M_B\sim 500$ GeV, and critical coupling.
The effects of new physics from topcolor at this level
are probably already ruled out, depending upon the
mass of the top quark.  Certainly, a more definitive
situation will exist at the end of run I(b) at the Tevatron.

Unfortunately, we cannot presently
argue that the experiments are any more
than marginally inconsistent with the standard model. The
standard model may end up prevailing in these observables.
However, our toy models illustrate the importance of refining
these measurements.   Moreover, we see that establishing the
validity of the standard model at the $Z$--pole is far from
establishing its validity at $\sim 500$ GeV at which
top production at the Tevatron occurs.  In the long run the possibility
of new physics beyond the standard model beginning to show up
in these observables is real, and may represent our first excursion into
the new territory of electroweak symmetry breaking.

\vskip .2in
\noindent
{\bf Acknowledgements}
\vskip .2in
\noindent
One of us (C.T.H.) wishes thank K. Sliwa for urging us to
look at this issue, and S. Parke for emphasizing out the
possibility of corollary phenomena in $e^+e^-\rightarrow \overline{b}b$
collisions at high $Q^2$ \cite{HP2}.
This work was performed at the Fermi National Accelerator Laboratory,
which is operated by Universities Research Association, Inc., under
contract DE-AC02-76CHO3000 with the U.S. Department of Energy.
XZ is grateful to the Theory Group
at FermiLab for hospitality. XZ was supported in part by the
Office of High Energy and Nuclear Physics of the U.S.
 Department of Energy (Grant No. DE-FG02-94ER40817 ).

\newpage
\vskip 0.2in
\begin{center}
\begin{tabular}{|| l | c | c | c | c ||}
\hline
 $ \Lambda $ GeV & $\delta\Gamma/
\Gamma_{b {\overline{b}}}$  & $R_b$ & $R_b/R_{b\;SM}$ &
$\sigma_{t\overline{t}}/\sigma_{SM}$ \\ \hline
  $ 300 $ &  $ 0.09803 $ &  $ 0.2368 $ &  $ 1.09803 $  & $\sim 6$ \\ \hline
  $ 600 $ &  $ 0.03874 $ &  $ 0.2241 $ &  $ 1.03875 $ & $2.4$ \\ \hline
  $ 900 $ &  $ 0.02092 $ &  $ 0.2202 $ &  $ 1.02092 $  & $1.3$ \\ \hline
  $ 1200 $ &  $ 0.01324 $ &  $ 0.2186 $ &  $ 1.01325 $ & $1.05$ \\ \hline
 \hline\hline
\end{tabular}
\end{center}
\begin{quote}
\vspace{.2in}
Table I: $R_b$ for various values of the cutoff
$\Lambda$ in the color singlet contact term case. $m_t=174$ GeV is
assumed, and we take the standard model
result to be $0.2157$, and thus $R_b/R_{b\;SM} = 1.0162\pm .0083$.
In the last column we give the ratio of the top pair production
cross-section to the standard model result,
where we also
assume the light fermions participate in four--fermion
operators as in  \cite{HP}.
Since this assumes that the new physics acts universally on the light
quarks, $u$ and $d$, it is in a sense an upper limit on
the effects on top production.
We see that saturating the
$R_b/R_{b\;SM}$ excess leads to a $\lta 30\%$ excess in top
production, while significantly larger top cross-sections are
ruled out by the observed $R_b/R_{b\;SM}$ if the physics is described by color
singlet contact terms.
\end{quote}
\vskip 0.2in

\newpage

\vskip 0.2in
\begin{center}
\begin{tabular}{|| l | c | c | c | c ||}
\hline
$N_cg_3^2\cot^2\theta/8\pi$  & $ M_V $ GeV & $\delta\Gamma/
\Gamma_{b {\overline{b}}}$  & $R_b$ & $R_b/R_{b\;SM}$ \\ \hline \hline
$ .5 $ &  $ 400 $ &  $ 0.007227 $ &  $ 0.2173 $ &  $ 1.0072 $   \\ \hline
$ 1 $ &  $ 400 $ &  $ 0.01445 $ &  $ 0.2188 $ &  $ 1.0145 $   \\ \hline
$ 1.5 $ &  $ 400 $ &  $ 0.02168 $ &  $ 0.2204 $ &  $ 1.0217 $   \\ \hline
$ 2 $ &  $ 400 $ &  $ 0.08909 $ &  $ 0.2219 $ &  $ 1.0289 $   \\ \hline
 \hline\hline
$ .5 $ &  $ 600 $ &  $ 0.00409 $ &  $ 0.2166 $ &  $ 1.0041 $   \\ \hline
$ 1 $ &  $ 600 $ &  $ 0.00818 $ &  $ 0.2175 $ &  $ 1.0089 $   \\ \hline
$ 1.5 $ &  $ 600 $ &  $ 0.01227 $ &  $ 0.2183 $ &  $ 1.0122 $   \\ \hline
$ 2 $ &  $ 600 $ &  $ 0.01636 $ &  $ 0.2192 $ &  $ 1.0164 $   \\ \hline
 \hline\hline
$ .5 $ &  $ 800 $ &  $ 0.002653 $ &  $ 0.2163 $ &  $ 1.0027 $   \\ \hline
$ 1 $ &  $ 800 $ &  $ 0.005305 $ &  $ 0.2168 $ &  $ 1.0053 $   \\ \hline
$ 1.5 $ &  $ 800 $ &  $ 0.007958 $ &  $ 0.2174 $ &  $ 1.0080 $   \\ \hline
$ 2 $ &  $ 800 $ &  $ 0.01061 $ &  $ 0.2180 $ &  $ 1.0106 $   \\ \hline
 \hline\hline
\end{tabular}
\end{center}
\begin{quote}
\vspace{.2in}
Table II: $R_b$ for various values of the parameters of the topcolor
model, where we take the standard model
result to be $0.2157$ and thus $R_b/R_{b\;SM} = 1.0162\pm .0083$. The value
$1.0$ for
$k=N_cg_3^2\cot^2\theta/8\pi$ corresponds to NJL critical coupling.
A $600$ ($800$) GeV coloron would
produce about a four--fold (two--fold) increase in the $t\overline{t}$
production
cross--section
relative to QCD (independently of $N_cg_3^2\cot^2\theta/8\pi$,
for a top mass of $175$ GeV) thus
top production is a more sensitive probe
of new physics than $Z\rightarrow b\overline{b}$
in the topcolor model.
\end{quote}
\vskip 0.2in

\newpage

\vskip 0.1in
\noindent
{\bf Figure Captions}
\vskip .2in
\noindent
Figures 1. Vertex correction to $Z b \overline{b}$.

\vskip .2in
\noindent
Figure 2. Radiative correction to the $Z b \overline{b}$ in the topcolor model.

\vskip .2in
\noindent
Figure 3(a). $R_b/R_{b\;SM}$ for various values of the parameters of the
topcolor
model, where the LEP result
$ = 1.0162\pm .0083$ is superimposed. The value $k=1.0$ where
$k= N_cg_3^2\cot^2\theta/8\pi$ corresponds to NJL critical coupling,
and we consider $k=0.5$, $k=1.0$ and $k=2.0$.

\vskip .2in
\noindent
Figure  3(b). The top production cross-section $\sigma_{t\overline{t}}$
normalized to the standard model result is given
for the topcolor model \cite{HP}; we take $\sigma_{t\overline{t}}(SM)
= 5.25 $ $pb$ for $m_t=175$ GeV, and we superimpose the CDF result
\cite{CDF}.  The Figures 3(a) and 3(b)
show that top cross--section measurements are
more restrictive than $R_b$ for the topcolor model; thus, to fit
the central value of the LEP $R_b$ requires an unacceptably large
top cross-section in the model.

\end{document}